\documentclass[twocolumn,aps,pra,superscriptaddress,showpacs]{revtex4}
\usepackage{graphicx}
\usepackage{amsmath}
\usepackage{amssymb}

\begin{document}

\title{Near-infrared single-photon spectroscopy of a whispering gallery mode resonator using energy-resolving transition edge sensors}

\author{Michael F\"{o}rtsch}
\affiliation{Max Planck Institute for the Science of Light, G\"{u}nther-Scharowsky-Str. 1, Bau 24, 91058, Erlangen, Germany}
\affiliation{Institut f\"{u}r Optik, Information und Photonik, University of Erlangen-Nuremberg, Staudtstr. 7/B2, 91058, Erlangen, Germany}
\affiliation{SAOT, School in Advanced Optical Technologies, Paul-Gordan-Str. 6, 91052 Erlangen}

\author{Thomas Gerrits}
\affiliation{National Institute of Standards and Technology, 325 Broadway, Boulder, CO 80305, USA}

\author{Martin J. Stevens}
\affiliation{National Institute of Standards and Technology, 325 Broadway, Boulder, CO 80305, USA}

\author{Dmitry Strekalov}
\affiliation{Max Planck Institute for the Science of Light, G\"{u}nther-Scharowsky-Str. 1, Bau 24, 91058, Erlangen, Germany}

\author{Gerhard Schunk}
\affiliation{Max Planck Institute for the Science of Light, G\"{u}nther-Scharowsky-Str. 1, Bau 24, 91058, Erlangen, Germany}
\affiliation{Institut f\"{u}r Optik, Information und Photonik, University of Erlangen-Nuremberg, Staudtstr. 7/B2, 91058, Erlangen, Germany}
\affiliation{SAOT, School in Advanced Optical Technologies, Paul-Gordan-Str. 6, 91052 Erlangen}

\author{Josef U. F\"{u}rst}
\affiliation{Max Planck Institute for the Science of Light, G\"{u}nther-Scharowsky-Str. 1, Bau 24, 91058, Erlangen, Germany}
\affiliation{Institut f\"{u}r Optik, Information und Photonik, University of Erlangen-Nuremberg, Staudtstr. 7/B2, 91058, Erlangen, Germany}

\author{Ulrich Vogl}
\affiliation{Max Planck Institute for the Science of Light, G\"{u}nther-Scharowsky-Str. 1, Bau 24, 91058, Erlangen, Germany}
\affiliation{Institut f\"{u}r Optik, Information und Photonik, University of Erlangen-Nuremberg, Staudtstr. 7/B2, 91058, Erlangen, Germany}

\author{Florian Sedlmeir}
\affiliation{Max Planck Institute for the Science of Light, G\"{u}nther-Scharowsky-Str. 1, Bau 24, 91058, Erlangen, Germany}
\affiliation{Institut f\"{u}r Optik, Information und Photonik, University of Erlangen-Nuremberg, Staudtstr. 7/B2, 91058, Erlangen, Germany}
\affiliation{SAOT, School in Advanced Optical Technologies, Paul-Gordan-Str. 6, 91052 Erlangen}

\author{Harald G. L. Schwefel}
\affiliation{Max Planck Institute for the Science of Light, G\"{u}nther-Scharowsky-Str. 1, Bau 24, 91058, Erlangen, Germany}
\affiliation{Institut f\"{u}r Optik, Information und Photonik, University of Erlangen-Nuremberg, Staudtstr. 7/B2, 91058, Erlangen, Germany}

\author{Gerd Leuchs}
\affiliation{Max Planck Institute for the Science of Light, G\"{u}nther-Scharowsky-Str. 1, Bau 24, 91058, Erlangen, Germany}
\affiliation{Institut f\"{u}r Optik, Information und Photonik, University of Erlangen-Nuremberg, Staudtstr. 7/B2, 91058, Erlangen, Germany}
\affiliation{Department of Physics, University of Ottawa, 550 Cumberland St,ON K1N 6N5, Canada}

\author{Sae Woo Nam}
\affiliation{National Institute of Standards and Technology, 325 Broadway, Boulder, CO 80305, USA}

\author{Christoph Marquardt}
\affiliation{Max Planck Institute for the Science of Light, G\"{u}nther-Scharowsky-Str. 1, Bau 24, 91058, Erlangen, Germany}
\affiliation{Institut f\"{u}r Optik, Information und Photonik, University of Erlangen-Nuremberg, Staudtstr. 7/B2, 91058, Erlangen, Germany}

\begin{abstract}
We demonstrate a method to perform spectroscopy of near-infrared single photons without the need of dispersive elements. This method is based on a photon energy resolving transition edge sensor and is applied for the characterization of widely wavelength tunable narrow-band single photons emitted from a crystalline whispering gallery mode resonator. We measure the emission wavelength of the generated signal and idler photons with an uncertainty of up to 2\,nm. \\
\textit{*Contribution of NIST, an agency of the U.S. Government, not subject to copyright}
\end{abstract}


\maketitle
\section{Introduction}
Near-infrared spectroscopy is an essential measurement technique in a large variety of research fields. Its applications range from physics over chemistry to biology and is also discussed as an application in forensics \cite{Stuart:2005}. Fundamentally, spectroscopy identifies the amount of radiation intensity as a function of wavelength, frequency, or energy.  Typically, spectroscopes consist of dispersive elements followed by either a movable slit and one photo-detector, or a detector array. Therefore, the wavelength range and the precision of this technology is mainly  limited by geometry. 

In this paper we demonstrate a method to directly measure the energy of the detected light at the single photon level without spectral discrimination. Due to the discrete nature of the photons, we can directly calculate the wavelength of the detected light. For the demonstration, we will calibrate the response of a transition edge sensor (TES) and perform spectroscopy on a single photon test source, which in our case will be a widely wavelength tunable crystalline whispering gallery mode resonator (WGMR). 
To the best of our knowledge, this is the first time that a TES is used for spectroscopic measurements in the infrared wavelength regime. 

The TES is a superconducting photon detector and operates as a microcalorimeter measuring the absorbed heat in the detection region. A number of experiments have demonstrated high energy resolution of a TES ranging from millimeter wave \cite{Bock:1995} to x-ray detection \cite{Irwin:1996}. The latter has shown remarkably good performance in x-ray spectroscopy \cite{Wollman:2000,Bennett:2012}. Here, we use a TES optimized for optical frequencies in the visible to near-infrared regime \cite{Cabrera:1998}. The TES is similar to devices with near-unity detection efficiency around 1550\,nm \cite{Lita:2008,Miller:2011}. Since the output signal of the TES is directly proportional to the amount of absorbed energy, this method enabled us to perform spectroscopic measurements at the single photon level without the use of additional wavelength dependent elements. 

WGMRs combine high quality factors (Q-factors) with small mode volumes over the whole transparency range of the used material \cite{Vahala:2003}. This makes them attractive for various experiments in nonlinear and quantum optics \cite{Spillane:2002,Ilchenko:2004,Fuerst:2010_1,Marquardt:2013,Strekalov:2013}. In the field of optical parametric oscillators (OPOs), for example, crystalline WGMRs have demonstrated generation of parametric light at remarkably low optical pump powers \cite{Fuerst:2010_2,Fuerst:2011}. Furthermore, these systems enable controlled generation of signal and idler beams over a large wavelength range without sacrificing the intrinsic narrow bandwidth \cite{Beckmann:2011,Werner:2012}. Recently we have successfully demonstrated an efficient narrow-band heralded single photon source, readily tunable in bandwidth and wavelength  based on a crystalline WGMR \cite{Fortsch:2013}. A similar source is used in this work to verify the performance of the TES-based spectroscopy. 

By combining the TES with the crystalline WGMR we measure a wavelength shift of the idler over a wavelength range 30\,nm with a uncertainty of 2\,nm. The corresponding signal photon shift is measured to be 15\,nm. Both results agree well with the theory of the temperature dependent wavelength shift of the source.

\section{Experimental Setup}
The starting point of our setup is a continuous wave (cw) Nd:YAG laser, emitting light at 1064\,nm. In order to simplify the future data analysis, a gated operation of the setup is preferred, where light from the WGMR is only impinging on the TES during predetermined time slots. Therefore, we generate temporal pulses from the initial cw light using a fast switching electro-optical modulator (EOM). The generated infrared pulses are subsequently frequency doubled using a periodically poled lithium niobate (PPLN) crystal. This cascaded pulse generation improves the limited extinction ratio of the EOM. 
\begin{figure}[t]
	\centering\includegraphics[width=0.48\textwidth]{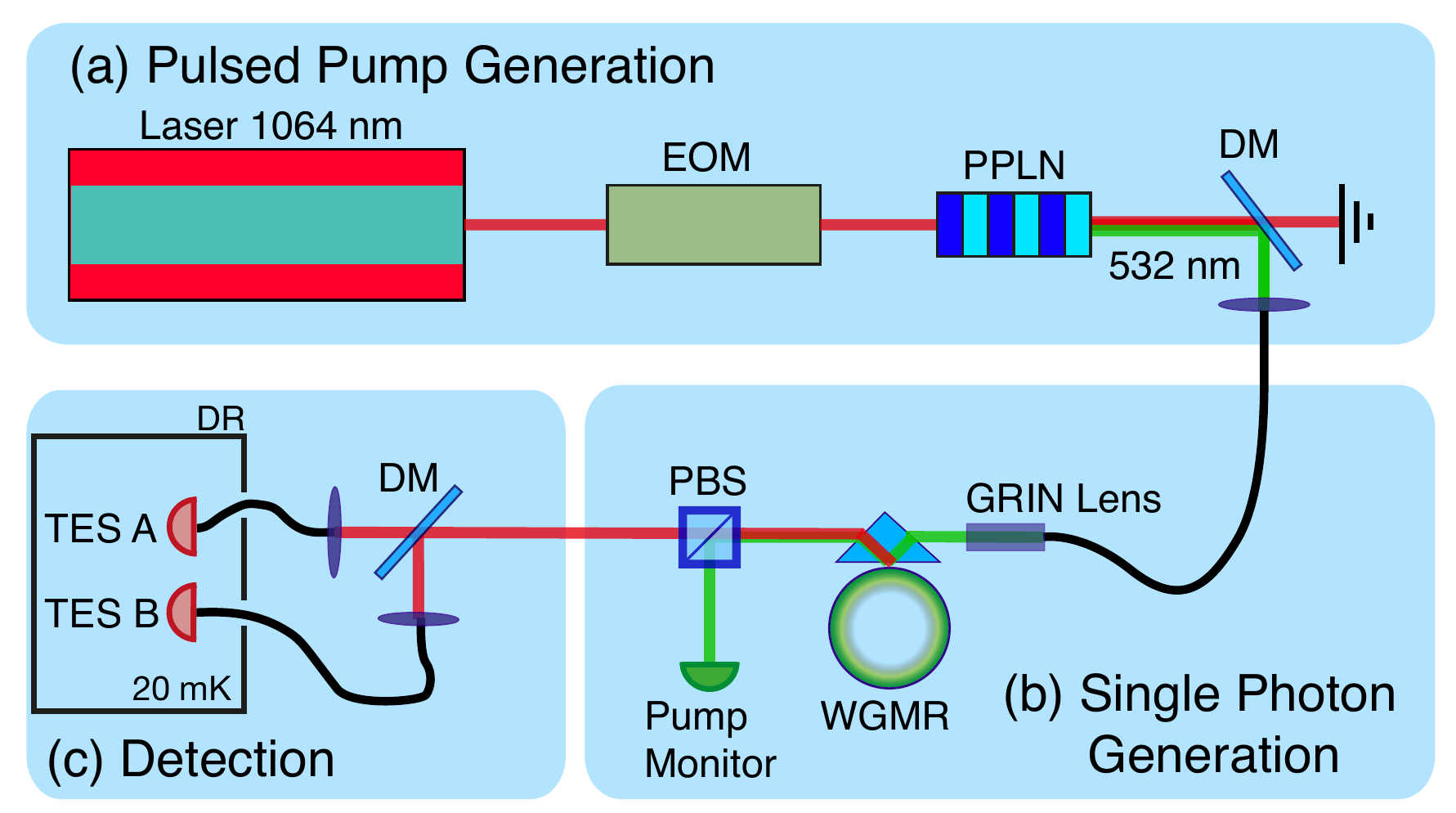}
	\caption{\label{fig:experiment}\textbf{Experimental Setup} (a) Preparing temporal shaped pulses at a wavelength of 532\,nm from a continuous wave Nd:YAG laser using an electro-optical modulator (EOM) and a periodical poled lithium niobate crystal (PPLn). (b) Generation of narrow-band single photon pairs with tunable wavelength utilizing a crystalline whispering gallery mode resonator (WGMR). (c) Detection of the corresponding energy of the emitted photons using a transition edge sensor (TES). DM: dichroic mirror; PBS: polarizing beam splitter}	
\end{figure}
With this technique we verified that no cw background at 532\,nm could be observed. The pulse length was chosen to be 100\,ns which corresponds approximately to twice the ring-down time of our WGMR. To minimize the effect of pulse pileup \cite{Rosenberg:2005} during the TES recovery time of $t_{\text{recover}}\approx 12\,\mu \text{s}$, the repetition rate was set to 80\,kHz. 

\begin{figure}[t]
	\centering\includegraphics[width=0.48\textwidth]{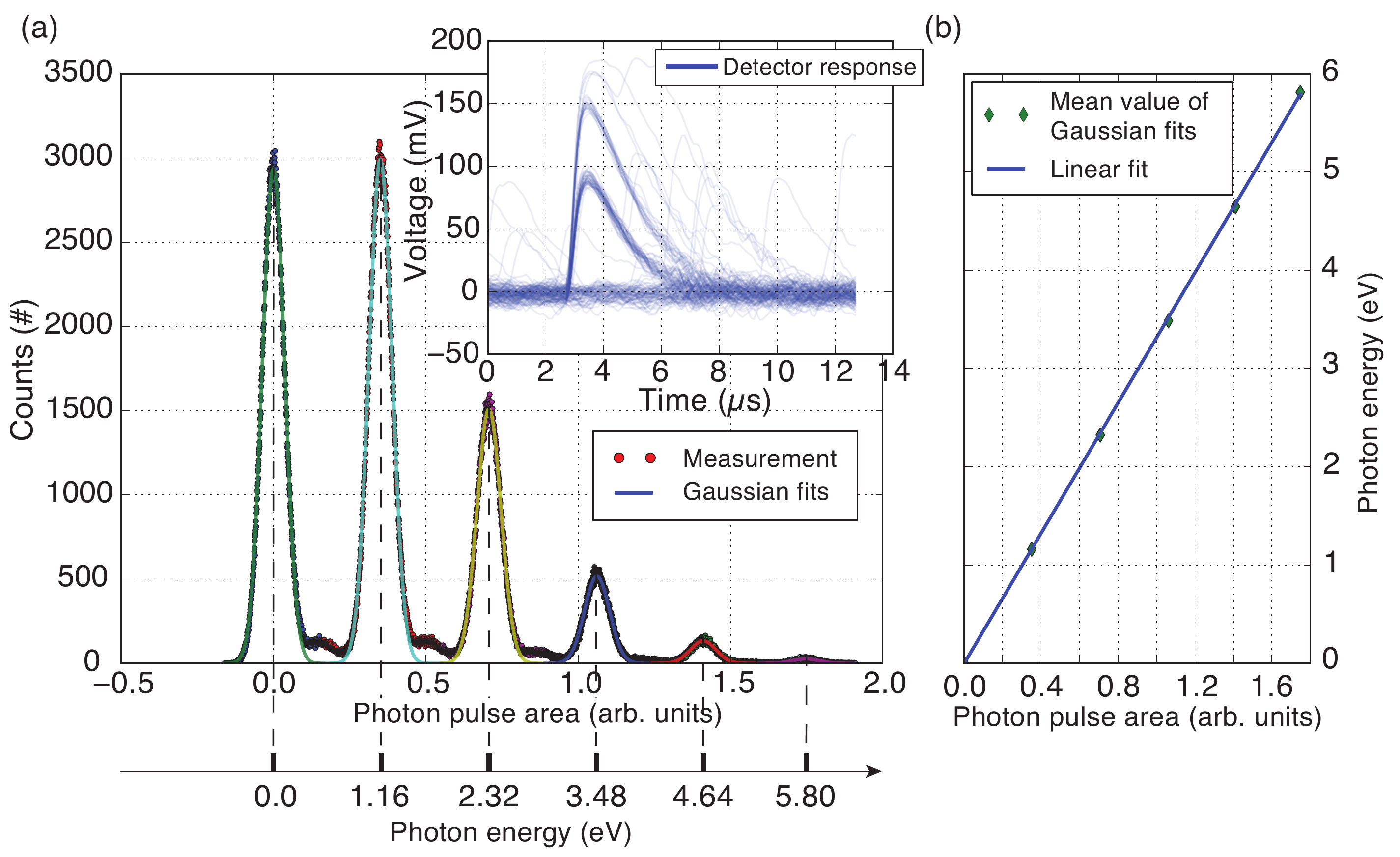}
	\caption{\label{fig:TES_calibration}\textbf{Detector calibration} (a) Photon pulse area distribution of a coherent state originating from the attenuated coherent diode laser source. The wavelength of the laser is 1062.9\,nm. The inset shows 1024 TES output traces. (b) The pulse area of the TES response pulse dependent on the absorbed energy. The green dots are the mean values from the Gaussian fits from (a).}
\end{figure}

The 532\,nm light is coupled to a single mode fiber and focused on the backside of a diamond prism using a GRIN lens. By controlling the distance between diamond prism and the WGMR, we control the optical tunneling between the beam's footprint at the prism and the resonator. 
The WGMR is manufactured from 5.8\% Mg-doped z-cut lithium niobate crystal (LiNbO$_{3}$) with a equatorial radius of $R\approx1.61$~mm and a polar radius of $r\approx0.4$~mm. The quality factor of the resonator at the pump wavelength of 532\,nm is $Q\approx 3\cdot10^{7}$.
By changing the stabilized temperature of the resonator, we control the phase-matching between the extraordinary polarized pump and the ordinary polarized downconverted signal and idler photons (natural Type I phase matching). The average pump power impinging on the coupling prism is $1.5\,\mu$W, which provides a sufficient signal to noise ratio on the pump monitor detector and simultaneously is still in the linear gain regime of the  downconversion process \cite{Fortsch:2013}. 
Signal and idler, as well as the residual pump light. exit the resonator via the diamond prism and are subsequently separated from each other using a polarizing beam splitter (PBS). The residual pump light is monitored using a PIN detector for locking the frequency of the pump laser to the WGMR. The non-degenerate signal and idler photons are further separated using a dichroic mirror and are directed to two fiber-coupled TESs.  
The TESs are operated at 20\,mK, which is far below the critical temperature $T_c$, using a commercial dilution refrigerator (DR). Both detectors are voltage biased and the electronic response of each is read out using a dc-superconducting interference device (SQUID), which is also placed inside the DR. After an additional $100\times$ amplification, each signal is recorded with a data-acquisition unit and post-processed to identify the energy of the detected photons. Examples of raw output pulses of the TES are shown in the inset of Fig.~\ref{fig:TES_calibration}(a). Since these TESs were optimized for a high photon number resolution, they have a low transition temperature, resulting in a relatively long recovery time of $t_{\text{recover}}\approx$12\,$\mu$s (compare inset Fig.~\ref{fig:TES_calibration}(a)).

\section{Energy calibration of the TES}
\label{sec:TES_calibration}
In order to determine the energy of the detected photon pulses, one has to calibrate the voltage response of each TES. This was done by using an attenuated coherent state from a diode laser source, which was electronically pulsed with 10\,ns pulse duration and a repetition rate of 35\,kHz. For each trigger pulse, the TES response is recorded for 13\,$\mu$s. The integral of this TES response function is associated with the energy of each detected photon pulse. The inset of Fig.~\ref{fig:TES_calibration}(a) shows typical TES responses for 1024 laser photon pulses. To improve the signal to noise ratio, each TES response is convoluted with a predetermined characteristic TES master pulse. The integral of this convolution is as well associated with the energy of each detected photon pulse. To quantify the amount of absorbed energy, in a first step we calculate the histogram of all photon pulse integrals (Fig.~\ref{fig:TES_calibration}(a)). The combination of the discrete nature of photons with the knowledge about the laser's emission wavelength of 1062.9\,nm implicates that each peak in Fig.~\ref{fig:TES_calibration}(a) is a multiple of 1.16\,eV. Therefore one can calibrate the histogram to the discrete amount of absorbed energy. This is done by fitting Gaussian distributions to the individual histogram peaks and correlating the mean value of the individual fits with the corresponding energy (as shown in Fig.~\ref{fig:TES_calibration}(a)). 
The mean values of the Gaussian fits are used to determine the scaling of the response integral depending on the incident energy  (Fig.~\ref{fig:TES_calibration}(b)).  This results in the characteristic detector energy response function and is used further on to connect the TES response to the amount of energy absorbed. 

\begin{figure}[t]
	\centering\includegraphics[width=0.48\textwidth]{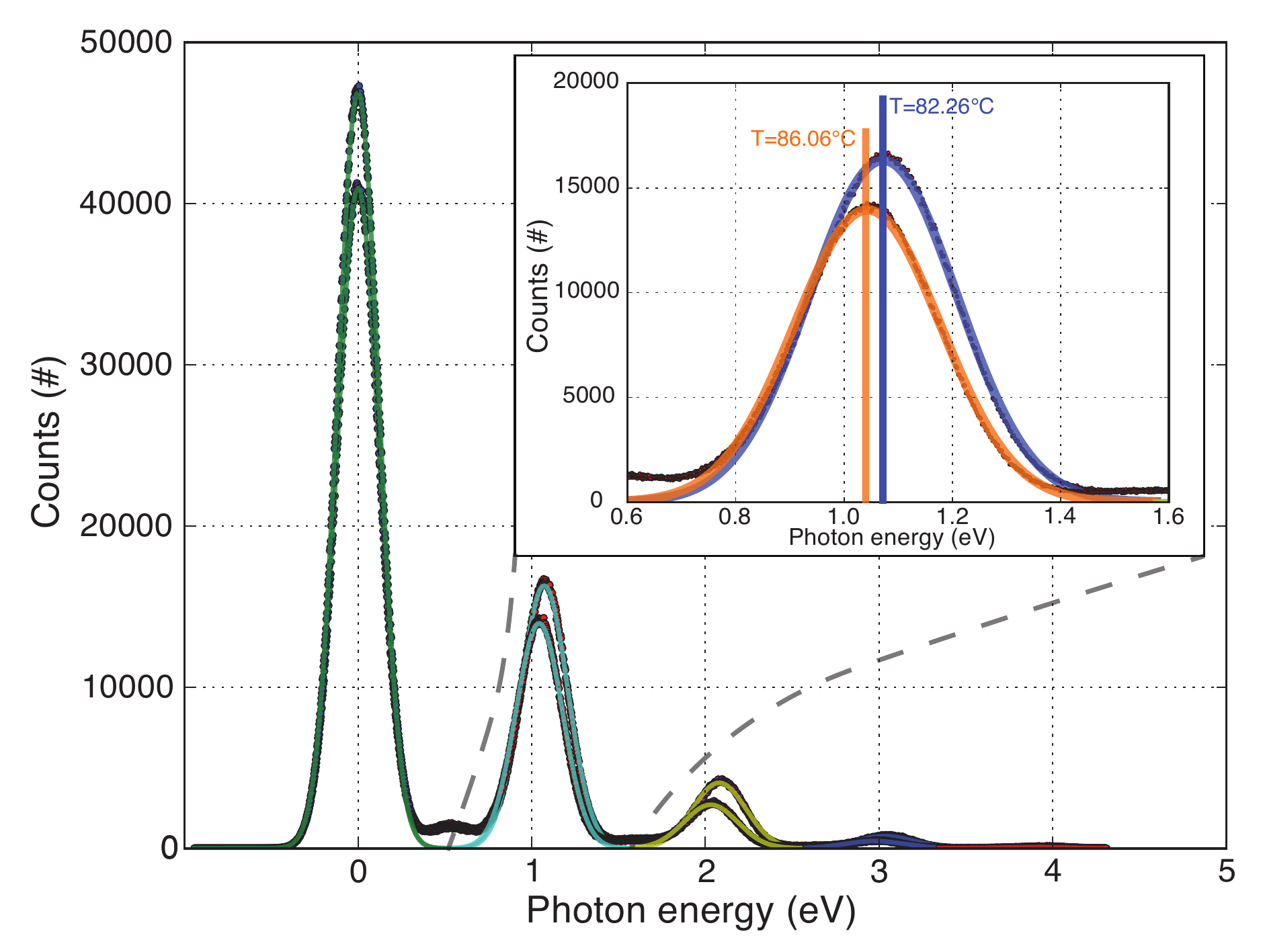}
	\caption{\label{fig:Tuning_histogram}\textbf{Photon pulse energy distributions} Photon pulse energy distributions measured in the idler arm for resonator temperatures of $82.26^{\circ}$C and  $86.06^{\circ}$C. The inset shows a close-up of the 1\,eV region.}
\end{figure}

\section{Spectroscopy of the generated pair photons}
\begin{figure}[t]
	\centering\includegraphics[width=0.48\textwidth]{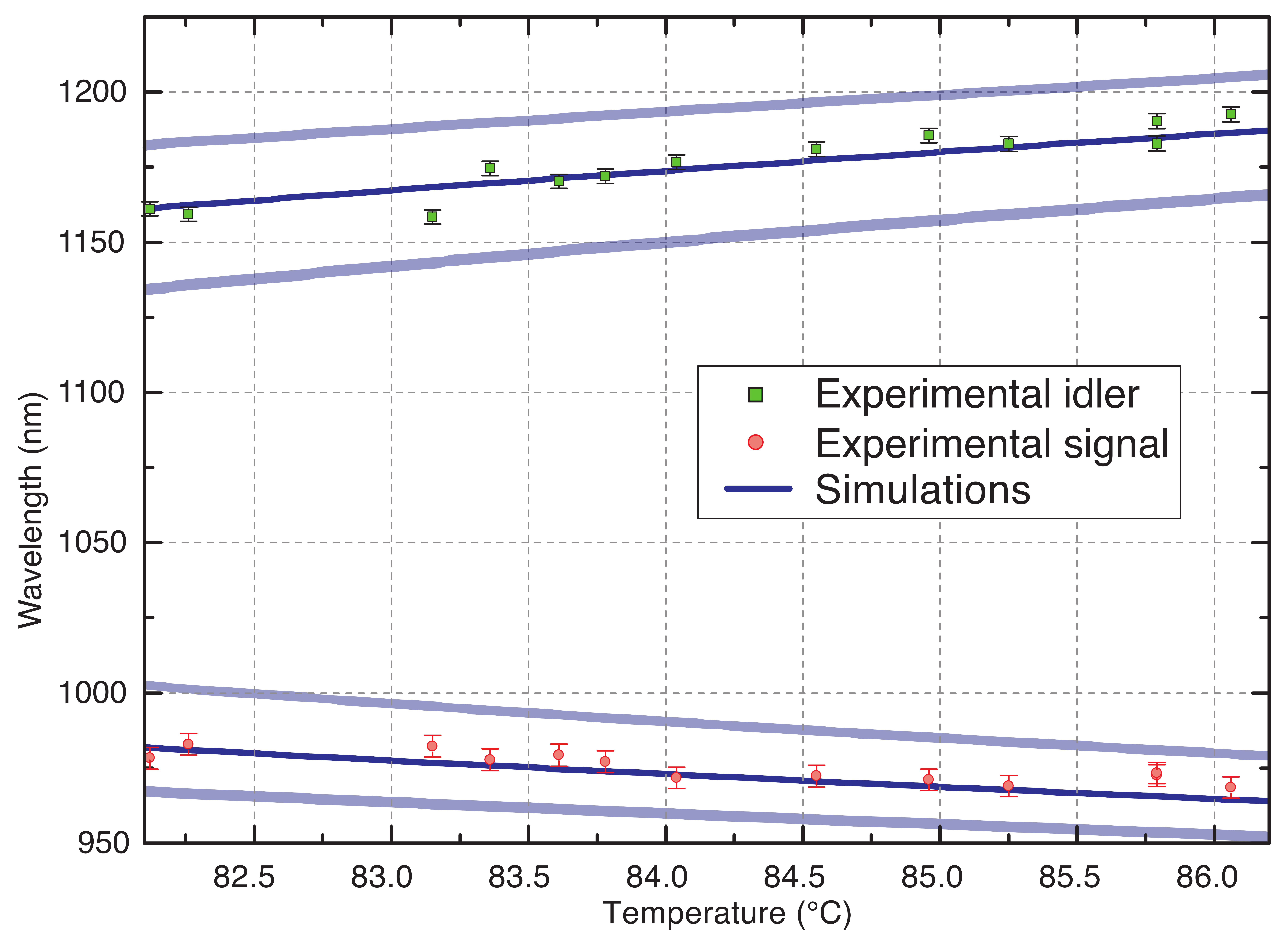}
	\caption{\label{fig:Wavelength_spectrum}\textbf{Wavelength spectrum} Signal and idler wavelengths measured for different phase-matching temperatures of the WGMR. The solid lines are the theoretical predictions of different phase-matching solutions \cite{Foertsch:2014}.}
\end{figure}
We characterize the wavelength of the emitted signal and idler photon pulses originating from the crystalline WGMR as a function of the phase-matching temperature. For a given phase-matching temperature we accumulate a photon pulse energy distribution for the signal and idler, respectively, using the characteristic energy response function of each detector. Fig.~\ref{fig:Tuning_histogram} illustrates two such energy histograms for two different phase-matching temperatures of the resonator, measured with the TES in the idler path. The inset of Fig.~\ref{fig:Tuning_histogram} shows a close-up of the region around 1\,eV, which is used to recalculate the wavelength of the detected photons. One can clearly see that the energy of the detected idler photons decreases when the resonator temperature increases. Due to the energy conservation of the SPDC process, this relation is reversed for the signal photons. In Fig.~\ref{fig:Wavelength_spectrum} the experimental result for the investigated temperature range for signal and idler is illustrated. By changing the resonator temperature over 4$^{\circ}$C we observe a wavelength detuning of nearly 30\,nm for the idler and 15\,nm for the signal photons. The error bars represent the one sigma interval of the mean-value determination using Gaussian fits to the photon pulse energy distributions as shown in Fig.~\ref{fig:Tuning_histogram}. For the fitting we used the Marquardt-Levenberg algorithm.  On average we realize a wavelength uncertainty of $\pm 2$\,nm for the idler and $\pm 4$\,nm for the signal. The main reason for this difference was an imbalanced fiber coupling efficiency for signal and idler, which resulted in a different number of total counts for signal and idler. 
For the theoretical prediction of the wavelength tuning in Fig.~\ref{fig:Wavelength_spectrum} (blue line), we calculate the wavelengths for signal $\lambda_s$ and idler $\lambda_i$ that fulfill the type I phase matching condition for our pump wavelength $\lambda_p$ \cite{Kozyreff:2008}. This condition strongly depends on the bulk refractive index, the shape of the resonator, the temperature, and the transverse mode structure \cite{Gorodetsky:2006}. As seen in Fig.~\ref{fig:Wavelength_spectrum} there exist various phase-matching solutions in the investigated temperature range that appear as clusters \cite{Foertsch:2014}. Our results are in good agreement with the numerical simulations for a fixed mode combination. This indicates that we successfully performed spectroscopy on the single photon level. 

\section{Summary}
We have used the energy resolution of superconducting transition edge sensors to resolve the wavelength of near-infrared single photons without the use of dispersive elements. The results were obtained by measuring the output wavelength of the signal and idler photons from a crystalline WGMR. We confirmed the theoretically expected wavelength shift of signal and idler photons based on a change of the phase-matching temperature. The presented measurement uncertainty of 2\,nm can be improved further by an optimized fiber coupling and by accumulating more photon counts. The absence of dispersive elements should allow us to extend this spectroscopy method to a wide wavelength range without increasing the measurement uncertainty. Moreover, by exploiting the photon number resolving ability of the TES, this spectroscopy method offers the possibility to characterize the quantum state of the detected photons \cite{Brida:2012} in combination with its wavelength. 

\end{document}